# Integer Reset Timed Automata: Clock Reduction and Determinizability

Lakshmi Manasa and Shankara Narayanan Krishna

Department of Computer Science & Engineering,
IIT Bombay, Powai, Mumbai-76, India.
{manasa,krishnas}@cse.iitb.ac.in

**Abstract.** In this paper, we propose a procedure that given an integer reset timed automaton (IRTA) $\mathcal{A}$, produces a language equivalent deterministic one clock IRTA $\mathcal{B}$ whose size is at most doubly exponential in the size of $\mathcal{A}$. We prove that this bound on the number of locations is tight. Further, if integer resets are used in stopwatch automata, a subclass of stopwatch automata which is closed under all boolean operations and for which reachability is decidable is obtained.

## 1 Introduction

It is well known that for timed automata [3], emptiness checking is PSPACE-complete. This has paved the way for using timed automata in the verification of real-timed systems - several algorithms, tools have been built. Even though emptiness checking is decidable, the questions of universality, inclusion are undecidable for non-deterministic timed automata with more than one clock. Further, timed automata cannot be determinized. Investigations have shown that even restricted classes like the one considered in [1] have undecidable universality. Some of the known classes where timed automata can be effectively determinized are event clock automata (ECA) [4] and integer reset timed automata (IRTA) [10]. [5] talks about a condition satisfying which, timed automata are determinizable. They give a procedure to obtain a language equivalent deterministic infinite timed tree corresponding to a timed automaton $\mathcal{A}$. The result is that $\mathcal{A}$ can be determinized if the number of clocks per node in this tree is bounded. ECA and IRTA fall into this category.

Integer reset timed automata were introduced in [10]. For a imed automaton $\mathcal{A}$ and IRTA $\mathcal{B}$, [10] and [11] decide the question "is $L(\mathcal{A}) \subseteq L(\mathcal{B})$" with non-primitive recursive complexity and EXPSPACE respectively. [12] gives a technique for obtaining a language equivalent determinized one clock IRTA $\mathcal{A}'$ from an IRTA $\mathcal{A}$, with a triply exponential blow up in the number of locations. Subsequently, [13] proposes a technique to obtain from an IRTA or $\epsilon$-IRTA $\mathcal{A}$, a one clock $\epsilon$-IRTA, with a doubly exponential blow up in the number of locations. The result in [13] cannot be considered an improvement over the one in [12] since the final IRTA obtained has $\epsilon$-moves (even when we start with an IRTA without $\epsilon$-moves). The determinization technique suggested in [5] applied to an IRTA $\mathcal{A}$, gives a deterministic timed automaton $\mathcal{B}$ (not an IRTA), whose size is doubly exponential in the size of $\mathcal{A}$, and which has $\leq c_m + 1$ clocks, where $c_m$ is the biggest constant used in the guards of $\mathcal{A}$.

As the main result of this paper, we show that starting with an IRTA $\mathcal{A}$, we can obtain a determinized one clock IRTA $\mathcal{B}$ whose size is doubly exponential in the size of $\mathcal{A}$. Comparing this result to the earlier works of [12], [13] and [5] we note the following.

- Our technique is extremely simple in comparison to the $\delta - \checkmark$ theory used in [12], [13]. [13] introduces $\epsilon$ moves in the one clock IRTA obtained even when the initial IRTA did not have any while [12] has a higher complexity.
- [5] gives rise to a deterministic timed automaton with $c_m + 1$ clocks, while we obtain a deterministic one clock IRTA.
- Finally, we prove that the doubly exponential bound is tight. This has not been established in any of these earlier works.

## 2 Preliminaries

For any set $S$, $S^*$ ($S^\omega$) denotes the set of all finite (infinite) strings over $S$. $S^\infty = S^* \cup S^\omega$. We consider as time domain $\mathbb{T}$ the set $\mathbb{Q}^+$ or $\mathbb{R}^+$ of non-negative rationals or reals, and $\Sigma$ a finite set of actions. A time sequence over $\mathbb{T}$ is a finite (infinite) non-decreasing sequence $t = (t_i)_{i \geq 0}$; for simplicity $t_0$ is taken to be zero always. For $t \in \mathbb{T}$, $int(t)$ and $frac(t)$ represent its integral and fractional parts respectively. A timed word over $\Sigma$ is defined as $\rho = (\sigma, t)$, where $\sigma = (\sigma_i)_{i \geq 1}$ is a finite (infinite) sequence of symbols in $\Sigma$ and $t = (t_i)_{i \geq 1}$ is a finite (infinite) sequence in $\mathbb{T}^\infty$. A *timed language* $L$ is a set of timed words.

We consider a finite set of variables $X$ called *clocks*. A clock valuation over $X$ is a map $\nu : X \to \mathbb{T}$ mapping each clock $x \in X$ to a time value. $\nu(x)$ represents the value assigned to the clock $x$ by $\nu$. For $t \in \mathbb{T}$, the valuation $\nu + t$ is defined as $(\nu + t)(x) = \nu(x) + t, \forall x \in X$. The set of all clock valuations over $X$ is denoted by $\mathbb{T}^X$. For the set of clocks $X$, the set of constraints (guards) over $X$, denoted by $C(X)$ is given by $\varphi ::= x \sim c | \varphi \wedge \varphi | \varphi \vee \varphi$ where $c \in \mathbb{N}, \sim \in \{<, \leq, >, \geq, =, \neq\}$. Clock constraints are interpreted over clock valuations. The relation $\nu \models \varphi$ (valuation $\nu$ satisfies constraint $\varphi$) is defined as $\nu \models x \sim c$ if $\nu(x) \sim c$. Clock constraints allow us to test the values of clocks. In order to change these values, we use the notion of *resets*. A reset $\phi$ is a subset of $X$ which mentions which set of clocks are reset. $\nu' = \nu[\phi := 0]$ denotes $\nu'(z) = \nu(z)$ for all $z \in X \backslash \phi$ and $\nu'(y) = 0$ for all $y \in \phi$. The set of all possible resets is $2^X$, the set of all subsets of $X$.

**Timed Automata**: A *timed automaton* [3] is a tuple $\mathcal{A} = (L, L_0, \Sigma, X, E, F)$ where $L$ is a finite set of locations; $L_0 \subseteq L$ is a set of initial locations; $\Sigma$ is a finite set of symbols; $X$ is a finite set of clocks; $E \subseteq L \times L \times \Sigma \times C(X) \times 2^X$ is the set of transitions and $F \subseteq L$ is a set of final locations. $C(X)$ and $2^X$ are the set of clock constraints and clock resets as described above. An edge $e = (l, l', a, \varphi, \phi)$ represents a transition from $l$ to $l'$ on symbol $a$, with the valuation $\nu \in \mathbb{T}^X$ satisfying the guard $\varphi$, and $\phi$ gives the resets of certain clocks. For a location $l$ and valuation $\nu$, $(l, \nu)$ is called a state of $\mathcal{A}$.

A path is a finite (infinite) sequence of consecutive transitions. The path is said to be accepting if it starts in an initial location ($l_0 \in L_0$) and ends in a final location (or repeats a final location infinitely often). A run $r$ through a path from a valuation $\nu'_0$ (with $\nu'_0(x) = 0$ for all $x$) is a sequence $(l_0, \nu'_0) \xrightarrow{t_1} (l_0, \nu_1) \xrightarrow{(\sigma_1, \varphi_1, \phi_1)} (l_1, \nu'_1) \xrightarrow{t_2} (l_1, \nu_2) \xrightarrow{(\sigma_2, \varphi_2, \phi_2)} (l_2, \nu'_2) \cdots (l_n, \nu'_n)$. Note that $\nu_i = \nu'_{i-1} + (t_i - t_{i-1}), \nu_i \models \varphi_i$, and that $\nu'_i = \nu_i[\phi_i := 0], i \geq 1$. The timed word corresponding to $r$ is $\rho = (\sigma_1, t_1)(\sigma_2, t_2) \cdots (\sigma_n, t_n)$. A timed word $\rho$ is accepted by $\mathcal{A}$ iff there exists an accepting run (through an accepting path) over $\mathcal{A}$, the word corresponding to which is $\rho$. The timed language $L(\mathcal{A})$ accepted by $\mathcal{A}$ is defined as the set of all timed words accepted by $\mathcal{A}$. In the following sections, we look at finite timed words.

**Region Automata**: Given a set $X$ of clocks, let $\mathcal{R}$ be a partitioning of $\mathbb{T}^X$. Each partition contains a set (possibly infinite) of clock valuations. Given $\alpha \in \mathcal{R}$, the successors of $\alpha$ represented by $Succ(\alpha)$ are defined as $\alpha' \in Succ(\alpha)$ if $\exists \nu \in \alpha, \exists t \in \mathbb{T}$ such that $\nu + t \in \alpha'$. The partition $\mathcal{R}$ is said to be a *set of regions* iff $\alpha' \in Succ(\alpha) \iff \forall \nu \in \alpha, \exists t \in \mathbb{T}$ such that $\nu + t \in \alpha'$. A set of regions is consistent with time elapse if two valuations which are equivalent (within the same partition) stay equivalent with time elapse. A region $\alpha \in \mathcal{R}$ is said to satisfy a clock constraint $\varphi \in C(X)$ denoted as $\alpha \models \varphi$, if $\forall \nu \in \alpha, \nu \models \varphi$. A clock reset $\phi \in 2^X$ maps a region $\alpha$ to a region $\alpha[\phi := 0] = \alpha'$ such that $\alpha' \cap \{\nu[\phi := 0]\} \neq \emptyset$ for some $\nu \in \alpha$. A set of regions $\mathcal{R}$ is said to be *compatible* with a set of clock constraints $C(X)$ iff $\forall \varphi \in C(X)$ and $\forall \alpha \in \mathcal{R}$ exactly one of the following holds (a) $\alpha \models \varphi$ or (b) $\alpha \models \neg \varphi$. A set of regions $\mathcal{R}$ is said to be *compatible* with a set of clock resets $2^X$ iff $\alpha' = \alpha[\phi := 0] \Rightarrow \forall \nu \in \alpha, \exists \nu' \in \alpha'$ such that $\nu' = \nu[\phi := 0]$.

Given a timed automaton $\mathcal{A}$, and a set of regions $\mathcal{R}$ compatible with $C(X)$ and $2^X$, the *region automaton* $\mathcal{R}(\mathcal{A}) = (Q, Q_0, \Sigma, E', F')$ is defined as follows: $Q = L \times \mathcal{R}$ the set of locations; $Q_0 = L_0 \times \{\alpha_0\}$ ($\alpha_0$ is the region where $\nu(x) = 0$ for all $x \in X$), the set of initial locations; $F' = F \times \mathcal{R} \subseteq Q$ the set of final locations; $E' \subseteq (Q \times \Sigma \times Q)$ is the set of edges. $(l, \alpha) \xrightarrow{a} (l', \alpha')$ is an edge in $E'$ if $\exists \alpha'' \in \mathcal{R}$ and a transition $(l, l', a, \varphi, \phi) \in E$ such that (a) $\alpha'' \in Succ(\alpha)$, (b) $\alpha'' \models \varphi$ and (c) $\alpha' = \alpha''[\phi := 0]$. The region automaton [3] is an abstraction of the timed automaton accepting $Untime(L(\mathcal{A}))$.

**Theorem 1.** *Let $\mathcal{A}$ be a timed automaton. Then the problem of checking emptiness of $L(\mathcal{A})$ is decidable. [3]*

## 2.1 Integer Reset Timed Automata

An *integer reset timed automaton* (IRTA) [10] is a timed automaton $\mathcal{A} = (L, L_0, \Sigma, X, E, F)$ with the restriction that for every $e = (l, l', a, \varphi, \phi) \in E$, if $\phi \neq \emptyset$ then $\varphi$ consists of atleast one atomic clock constraint $x = c$ for some $x \in X, c \in \mathbb{N}$. The clock constraint $x = c$ in the guard of a resetting transition ensures that all the resets happen at integer time units (see also Lemma 1). The timed automaton $\mathcal{A}$ shown in Figure 2.1 is an IRTA.

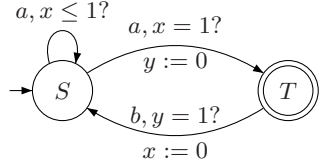

**Fig. 2.1.** IRTA $\mathcal{A}$.

**Lemma 1.** *[11] Let $\mathcal{A} = (L, L_0, \Sigma, X, E, F)$ be an IRTA and $\nu$ be a clock valuation in any given run in $\mathcal{A}$. Then $\forall x, y \in X$, $frac(\nu(x)) = frac(\nu(y))$.*

### 2.2 IRTA Regions

In this section, we look at the regions $\mathcal{R}$ of an IRTA. Given a set $X$ of clocks, let $\mathcal{R}$ be a finite partitioning of $\mathbb{T}^X$. The notions of successor of a region, compatibility with guards and compatibility with resets are same as mentioned earlier.

Let $c_m \in \mathbb{N}$ be the maximum constant occurring in the guards $C(X)$ of the IRTA $\mathcal{A}$. For the set of clocks $X$, define a set of intervals $\mathcal{I}$ as

$$\mathcal{I} = \{[c] | 0 \le c \le c_m\} \cup \{(c, c+1) | 0 \le c < c_m\} \cup \{(c_m, \infty)\}$$

We denote the clock interval of $t \in \mathbb{T}$ as $\langle t \rangle_{\mathcal{I}}$. For example, if $c_m = 2$, then $\langle 1 \rangle_{\mathcal{I}} = [1]$, $\langle 1.2 \rangle_{\mathcal{I}} = (1, 2)$ and $\langle 2.4 \rangle_{\mathcal{I}} = (2, \infty)$.

Let $\alpha$ be a tuple $((I_x)_{x \in X}, \prec)$ where (i) $I_x \in \mathcal{I}$ is the clock interval of $x \in X$, (ii) $\prec$ is a total preorder on $X_0 = \{x \in X \mid I_x \text{ is of the form } (c, c+1)\}$. The region associated with $\alpha$ is the set of valuations $\nu \in \mathbb{T}^X$ such that for all $x \in X$, $\nu(x) \in I_x$ and for all $x, y \in X_0$, $x \prec y$ iff $frac(\nu(x)) \le frac(\nu(y))$. Since the fractional parts of all clocks are same always (Lemma 1), we can drop the preorder $\prec$ and consider $\alpha$ to be $((I_x)_{x \in X})$. For $x \in X$, $\alpha(x) = I_x$. The set of all such tuples $\alpha$ partitions $\mathbb{T}^X$ and this is the set we consider to be $\mathcal{R}$. For a valuation $\nu$, the clock region it belongs to is denoted as $\langle \nu \rangle_{\mathcal{R}}$. For example, if $\nu(x) = 2.3, \nu(y) = 1.3, c_m = 3$, then $\langle \nu \rangle_{\mathcal{R}} = ((2, 3), (1, 2))$. We drop the subscripts for the notations $\langle t \rangle_{\mathcal{I}}$ and $\langle \nu \rangle_{\mathcal{R}}$ whenever they are clear from the context.

Consider the set of clock intervals $\mathcal{I}$ and the set of clock regions $\mathcal{R}$ defined for the set of clocks $X$ with the maximum clock constant being $c_m$. For two clock intervals $I_1, I_2 \in \mathcal{I}$, we define $I_1 + I_2$ as the clock interval $I \in \mathcal{I}$ such that $\forall t_1 \in I_1, \forall t_2 \in I_2, \exists t \in I$, such that $t = t_1 + t_2$. For a clock region $\alpha = (\{I_x\}_{x \in X}) \in \mathcal{R}$ and a clock interval $I \in \mathcal{I}$, we define $\alpha + I$ as the region $(\{I_x + I\}_{x \in X})$.

**Definition 1.** *Two timed words $\rho = (\sigma_1, t_1)(\sigma_2, t_2) \cdots (\sigma_n, t_n)$ and $\rho' = (\sigma'_1, t'_1)(\sigma'_2, t'_2) \cdots (\sigma'_n, t'_n)$ are said to be equivalent denoted by $\rho \cong \rho'$ iff for all $i$ the following holds (1) $\sigma_i = \sigma'_i$ and (2) $int(t_i) = int(t'_i)$, $frac(t_i) = 0$ iff $frac(t'_i) = 0$.*

**Lemma 2.** *If $\mathcal{A}$ is an IRTA and $\rho \cong \rho'$ then, $\rho \in L(\mathcal{A})$ iff $\rho' \in L(\mathcal{A})$ [11].*

Consider the timed automaton $\mathcal{A}$ in figure 2.2 and two timed words $\rho_1 = (a, 0.5)(c, 1.5)$ and $\rho_2 = (a, 0.5)(c, 1.4)$. $\rho_1 \cong \rho_2$. However $\rho_1 \in L(\mathcal{A})$ while $\rho_2 \notin L(\mathcal{A})$. This shows that lemma 2 need not hold for a timed automaton which is not an IRTA.

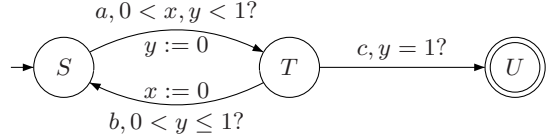

**Fig. 2.2.** Timed automaton $\mathcal{A}$ which is not an IRTA.

*Integral, Non-integral, Saturated region*: Let $\alpha = ((I_x)_{x \in X}) \in \mathcal{R}$ and let $X_m \subseteq X$ be such that $\forall x \in X_m, I_x = (c_m, \infty)$. (i) $\alpha$ is said to be saturated if $X_m = X$, (ii) $\alpha$ is said to be integral if $\forall x \in X \setminus X_m$, with $X_m \subset X$, $I_x$ is of the form $[c]$, and (iii) $\alpha$ is said to be non-integral if $\forall x \in X \setminus X_m$, with $X_m \subset X$, $I_x$ is of the form $(c, c+1)$. If $\mathcal{A}$ is an IRTA, and $\alpha$ is a region of $\mathcal{A}$, then $\alpha$ can be classfied as one of integral, non-integral or saturated region (Lemma 1 implies this). The union of the integral, saturated regions is denoted by $\mathcal{R}_I$. Following [7], we have

**Lemma 3.** *The set $\mathcal{R}$ of IRTA regions forms a set of regions. $\mathcal{R}$ is compatible with the clock constraints $C(X)$ and with the set $2^X$ of clock resets.*

## 3 Clock reduction and determinization of IRTA

In this section, we give a technique to obtain given an IRTA $\mathcal{A}$ with $k \geq 1$ clocks, an IRTA $\mathcal{A}^1$ with one clock $n$. As the constraints in $\mathcal{A}^1$ are over a single clock $n$, we can consider each constraint to be a disjunction of clock intervals from the set $\mathcal{I}$. For example, a constraint $n \leq 2 \wedge n \geq 1$ on a transition from $s$ to $t$ can be expressed as three transitions from $s$ to $t$ on $n \in [1]$, $n \in (1, 2)$ and $n \in [2]$ respectively. Let $c_m$ be the maximum constant used in the guards of $\mathcal{A}$. Given a clock region $\alpha$ of $\mathcal{A}$ and a constraint $\varphi^1$ of the form $n \in I_n$, $\alpha + \varphi^1$ consists of valuations obtained by adding $I_n$ to each interval $I_x$ in $\alpha$ (as defined in Section 2). For example, if $\alpha = (1 < x < 2, 0 < y < 1)$ and $\varphi^1 = n \in [1]$, then $\alpha + \varphi^1$ consists of the valuations $(2 < x < 3, 1 < y < 2)$. For a constraint $\varphi$ over $X$, the relation $\alpha + \varphi^1 \models \varphi$ iff $\nu \models \varphi$ for all $\nu \in \alpha + \varphi^1$. So, if $\varphi$ is $y > 2$, then in the example above, $\alpha + \varphi^1 \not\models \varphi$. However, $\alpha + \varphi^1 \models y > 1$. This notation will be used in the following construction.

### 3.1 Clock reduction

Given an IRTA $\mathcal{A} = (L, L_0, \Sigma, X, E, F)$ construct a one clock IRTA $\mathcal{A}^1 = (L^1, L_0^1, \Sigma, \{n\}, E^1, F^1)$ as follows:

- $L^1 \subseteq L \times \mathcal{R}_I$, $\mathcal{R}_I$ is the set of integral and saturated regions;
- $L_0^1 = L_0 \times \{\alpha_0\}$ where $\alpha_0 = ([0], [0], \cdots [0])$;
- $F^1 \subseteq F \times \mathcal{R}_I$;
- $E^1 \subseteq L^1 \times \Sigma \times \mathcal{I} \times 2^{\{n\}} \times L^1$ is the set of transitions. A transition $(l, \alpha) \xrightarrow{a, \varphi^1, \phi^1}$ $(l', \alpha')$ is defined iff there exists a transition $l \xrightarrow{a, \varphi, \phi} l'$ in $E$ such that
  - $\alpha + \varphi^1 \models \varphi$,
  - $\alpha' = (\alpha + \varphi^1)[\phi := 0]$ if $\phi \neq \emptyset$; $\alpha' = \alpha$ if $\phi = \emptyset$,
  - $\phi^1 = \{n\}$ iff $\phi$ is non-empty.

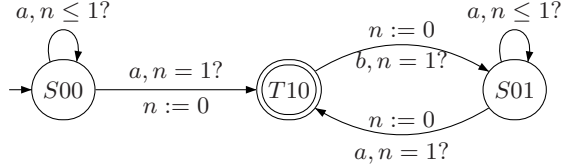

**Fig. 3.1.** One clock IRTA $\mathcal{A}^1$ corresponding to the IRTA $\mathcal{A}$ in Figure 2.1. $S01$ represents the location $S, (x = 0, y = 1)$

By construction, the region component $\alpha$ in the locations $(l, \alpha)$ of $\mathcal{A}^1$ is updated only whenever a reset happens in $\mathcal{A}$. Since resets happen only at integer time units, the region components are always integral. A reset in $\mathcal{A}$ results in resetting $n$ in $\mathcal{A}^1$; the value of $n$ is otherwise the time elapsed between two resets. Next, we prove that $\mathcal{A}$ and $\mathcal{A}^1$ accept the same timed language.

In the following proof, we represent a state $((l, \alpha), \mu)$ of $\mathcal{A}^1$ as $(l, \alpha, \mu)$ and use the notation $\nu = \alpha + \mu$ to represent that for all $x \in X, \nu(x) = c_x + \mu(n)$ where $[c_x] = \alpha(x)$.

**Theorem 2.** *Let $\mathcal{A}$ be an IRTA and let $\mathcal{A}^1$ be the one clock IRTA obtained using the above construction. Then $L(\mathcal{A}) = L(\mathcal{A}^1)$.*

*Proof.* $L(\mathcal{A}) \subseteq L(\mathcal{A}^1)$: Consider a run $(l_0, \nu_0') \xrightarrow{t_1} (l_0, \nu_1) \xrightarrow{\sigma_1, \varphi_1, \phi_1} (l_1, \nu_1')$ in $\mathcal{A}$ of length one. By construction of $\mathcal{A}^1$, there is a run $(l_0, \alpha_0, \mu_0') \xrightarrow{t_1}$ $(l_0, \alpha_0, \mu_1) \xrightarrow{\sigma_1, \varphi_1^1, \phi_1^1} (l_1, \alpha_1, \mu_1')$ where $\mu_0' = 0, \alpha_0 + \varphi_1^1 \models \varphi_1$. $\varphi_1^1$ is $n \in \langle t_1 \rangle$. Also, $\nu_0' = \alpha_0 + \mu_0'$, $\nu_1 = \alpha_0 + \mu_1$, $\nu_1' = \alpha_1 + \mu_1'$ irrespective of $\phi_1$.

Assume the result for all runs of length $< m$. Consider a run of $\mathcal{A}$ of length $m$. Let $(l_0, \nu_0') \xrightarrow{t_1} (l_0, \nu_1) \xrightarrow{\sigma_1, \varphi_1, \phi_1} (l_1, \nu_1') \ldots \xrightarrow{t_{m-1}} (l_{m-2}, \nu_{m-1}) \xrightarrow{\sigma_{m-1}, \varphi_{m-1}, \phi_{m-1}}$ $(l_{m-1}, \nu_{m-1}') \xrightarrow{t_m} (l_{m-1}, \nu_m) \xrightarrow{\sigma_m, \varphi_m, \phi_m} (l_m, \nu_m')$ be a run in $\mathcal{A}$ corresponding to $(\sigma_1, t_1) \ldots (\sigma_m, t_m)$. Consider the subrun $(l_0, \nu_0') \xrightarrow{t_1} (l_0, \nu_1) \xrightarrow{\sigma_1, \varphi_1, \phi_1}$ $(l_1, \nu_1') \ldots \xrightarrow{\sigma_{m-1}, \varphi_{m-1}, \phi_{m-1}} (l_{m-1}, \nu_{m-1}')$. By induction hypothesis, we can obtain a run of length $m - 1$ in $\mathcal{A}^1$ which ends in $(l_{m-1}, \alpha_{m-1}, \mu_{m-1}')$. The subrun in $\mathcal{A}$ extends as $(l_{m-1}, \nu_{m-1}') \xrightarrow{t_m} (l_{m-1}, \nu_m) \xrightarrow{\sigma_m, \varphi_m, \phi_m} (l_m, \nu_m')$. We know that $\nu_m \models \varphi_m$ and $\nu_m = \nu_{m-1}' + (t_m - t_{m-1})$. From induction hypothesis, we also know that $\nu_{m-1}' = \alpha_{m-1} + \mu_{m-1}'$. Hence there should exist edges $(l_{m-1}, \alpha_{m-1}, \mu_{m-1}') \xrightarrow{t_m} (l_{m-1}, \alpha_{m-1}, \mu_m) \xrightarrow{\sigma_m, \varphi_m^1, \phi_m^1} (l_m, \alpha_m, \mu_m')$. Since

$\nu_m = \alpha_{m-1} + \mu_m \models \varphi_m$, and $\alpha_{m-1} + \varphi_m^1 \models \varphi_m$, we have $\varphi_m^1 = n \in \langle \mu_m(n) \rangle$, and $\nu'_m = \alpha_m + \mu'_m$. Clearly, $(\sigma_1, t_1) \ldots (\sigma_m, t_m)$ is in $L(\mathcal{A}^1)$ whenever it is in $L(\mathcal{A})$. See Appendix A for an example.
$L(\mathcal{A}^1) \subseteq L(\mathcal{A})$: The above argument can be traced backward to argue this. □

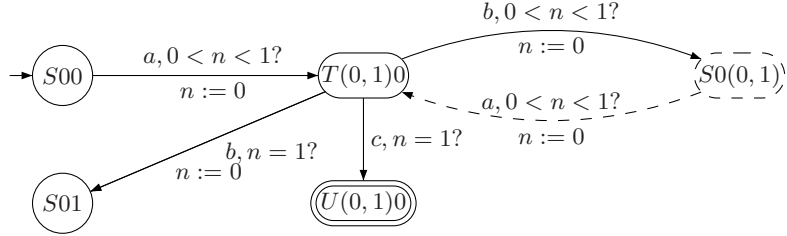

**Fig. 3.2.** One clock automaton $\mathcal{A}^1$ for the timed automaton in the Figure 2.2. $T(0,1)0$ represents the location $T, (0 < x < 1, y = 0)$.

However, it must be noted that this technique works because $\mathcal{A}$ is an IRTA. The fact that resets happen at globally integral times has helped us retain in $n$ the time elapsed between two resets. See the automaton $\mathcal{A}^1$ in Figure 3.2 which is obtained by applying the above technique to the timed automaton $\mathcal{A}$ in Figure 2.2. In the Figure 3.2, consider the location $[S0(0,1)]$ and the outgoing edge $[S0(0,1)] \overset{a,0<n<1?n:=0}{\longrightarrow} [T(0,1)0]$ (dotted in the figure). This edge corresponds to the edge $S \overset{a,0<x,y<1,y:=0}{\longrightarrow} T$ in $\mathcal{A}$ of Figure 2.2. Here the requirement $\alpha + \varphi^1 \models \varphi$ of the construction does not hold - not all valuations in $(x = 0, 0 < y < 1) + (0, 1)$ satisfy the constraint $0 < x, y < 1$. To satisfy $0 < x, y < 1$, we need to know the exact value of $y$. This can be achieved by (1) having a fresh clock containing value of $y$ or (2) remember the value of $y$ in the location. Option (2) would give rise to infinitely many locations in place of $[S0(0,1)]$. To sum up, the technique described above to reduce the number of clocks to one does not work for timed automata in general. It is worthwhile to mention Finkel's result [9] that the problem of the minimization of the number of clocks of a timed automaton is undecidable.

**Complexity** The definition of $\mathcal{A}^1$ shows that the number of locations is at most $|L| \times |\mathcal{R}_I| = |L| \times [c_m + 2]^{|X|}$. However, $E^1$ reveals that the region part of in $(l, \alpha)$ changes only if the corresponding edge in $\mathcal{A}$ resets at least one clock. Hence all the locations in $L^1$ have integral regions with at least one of the clocks having the interval $[0]$. Thus the total number of locations in $\mathcal{A}^1$ is $|L^1| \leq |L|.[(c_m + 2)^{|X|} - (c_m + 1)^{|X|}]$. Lemma 4 shows that this bound is indeed tight.

**Lemma 4.** *There is an IRTA $\mathcal{A}$ such that the smallest one clock IRTA $\mathcal{A}^1$ corresponding to it has exactly $|L|.[(c_m + 2)^{|X|} - (c_m + 1)^{|X|}]$ locations, where $L$ is the set of locations of $\mathcal{A}$, $X$ is the set of clocks of $\mathcal{A}$ and $c_m$ is the maximum constant used in the guards of $\mathcal{A}$.*

*Proof.* Consider the IRTA $\mathcal{A} = (L, L_0, \Sigma, X, E, F)$ in Figure 3.3 having two clocks. The one clock IRTA $\mathcal{A}^1$ in Figure 3.3 has exactly $|L|.[(c_m+2)^{|X|} - (c_m+1)^{|X|}]$ number of locations.

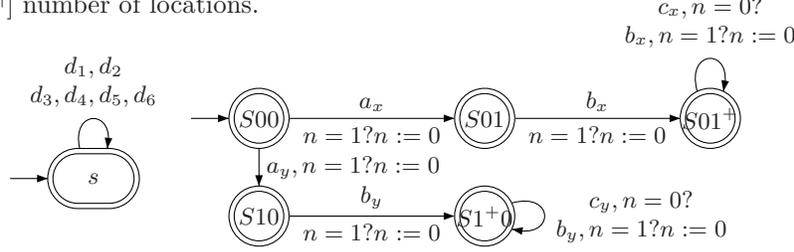

**Fig. 3.3.** Deterministic IRTA $\mathcal{A}$ and its one clock IRTA $\mathcal{A}'$. The symbols represent the following timed transitions : $d_1 ::= a_x, x = y = 1?x := 0$, $d_2 ::= a_y, x = y = 1?y := 0$, $d_3 ::= c_x, x = 0 \wedge y > 1?$, $d_4 ::= c_y, y = 0 \wedge x > 1?$, $d_5 ::= b_y, x > 1 \wedge y = 1?y := 0$, $d_6 ::= b_x, x = 1 \wedge y > 1?x := 0$. $1^+$ denotes all values $> 1$.

The language accepted by $\mathcal{A}$ is $L(\mathcal{A}) = \{(a_x, 1), (a_y, 1), (a_x, 1)(b_x, 2), (a_y, 1)(b_y, 2), (a_x, 1)(b_x, 2)(c_x, 2)(b_x, 3), (a_y, 1)(b_y, 2)(c_y, 2)(c_y, 2), \ldots\}$. Clearly, $untime(L(\mathcal{A})) = a_x(b_x c_x^*)^* + a_y(b_y c_y^*)^*$. It is easy to see that the minimal (deterministic, not complete) automaton $\mathcal{D}$ accepting $untime(L(\mathcal{A}))$ requires 5 locations (use the standard Myhill-Nerode argument). Decorating this with appropriate constraints (see below), we obtain a one clock IRTA $\mathcal{A}^1$ accepting $L(\mathcal{A})$.

To argue that $\mathcal{A}^1$ is the smallest one clock IRTA accepting $L(\mathcal{A})$ is easy: (1) To accept $(a_x, 1), (a_y, 1)$, we need two locations $s, t$ ($s$ is the initial location) with $s \xrightarrow{a_x, a_y, n=1?, n:=0} t$; (2) To accept $(a_x, 1)(b_x, 2)$, we reset the clock $n$ on the transition from $s$ to $t$ and add $t \xrightarrow{b_x, n=1?} s$. But this would mean accepting illegal words like $(a_y, 1)(b_x, 2), (a_y, 1)(b_x, 2)(a_x, 2)$ as well, hence we need to add new locations $u, v$ and replace $s \xrightarrow{a_y, n=1?n:=0} t$ with $s \xrightarrow{a_y, n=1?n:=0} u$ and replace $t \xrightarrow{b_x, n=1?} s$ with $t \xrightarrow{b_x, n=1?n:=0} v$; (3) After (2), to accept $(a_x, 1)(b_x, 2)(b_x, 3)\ldots(b_x, n)\ldots$, we need a loop on $b_x$ resetting $n$ every time $n = 1$. This is easily done by adding $v \xrightarrow{b_x, n=1?n:=0} v$. To incorporate any number of $c_x$'s without time elapse, we also add $v \xrightarrow{c_x, n=0?} v$. A similar argument will show that we need one more location $w$ to take care of $b_y, c_y$. It can be seen that what we obtain is precisely $\mathcal{A}^1$. □

### 3.2 Determinization

In this section, we give a technique to obtain from an IRTA $\mathcal{A}$, a one clock deterministic IRTA $\mathcal{A}^d$.

Given an IRTA $\mathcal{A} = (L, L_0, \Sigma, X, E, F)$, a language equivalent one clock deterministic IRTA $\mathcal{A}^d = (L^d, L_0^d, \Sigma, \{n\}, E^d, F^d)$ is constructed as follows:

- $L^d \subseteq 2^{L \times \mathcal{R}_I}$, where $\mathcal{R}_I$ is the set of integral and saturated regions;
- $L_0^d = \bigcup L_0 \times \{\alpha_0\}$ where $\alpha_0 = ([0], [0], \cdots [0])$;
- $F^d = \{A \in L^d \mid A \text{ contains some } (l, \alpha), l \in F\}$;
- $E^d \subseteq L^d \times \Sigma \times \mathcal{I} \times 2^{\{n\}} \times L^d$ is the set of transitions. Let $A = \{(l_1, \alpha_1), \ldots, (l_n, \alpha_n)\}$. A transition $A \xrightarrow{a, \varphi^d, \phi^d} B \in E^d$ iff
  - For each $(l_i, \alpha_i) \in A$, if there exists in $E$ an edge $l_i \xrightarrow{a, \varphi_i, \phi_i} l_i'$ such that $\alpha_i + \varphi^d \models \varphi_i$ then $(l_i', \alpha_i') \in B$,
  - $\phi^d = \{n\}$ iff $\phi_i \neq \emptyset$ for some $i \in \{1, 2, \ldots n\}$,
  - If $\phi^d = \{\}$ then $\alpha_i' = \alpha_i$ for all $i$. If $\phi^d = \{n\}$, then $\alpha_i' = \alpha_i + \varphi_d[\phi_i := 0]$ when $\phi_i \neq \{\}$ and $\alpha_i' = \alpha_i + \varphi_d$ when $\phi_i = \{\}$.

Figure 3.4 gives the deterministic one clock IRTA $\mathcal{A}^d$ obtained for the IRTA $\mathcal{A}$ in Figure 2.1. Note that the same can be achieved by determinising $\mathcal{A}^1$ (of Figure 3.1) in the same way [see Appendix B].

The technique outlined above is very similar to the one studied in the Section 3.1 except that it performs subset construction along with clock reduction. For example consider the automata $\mathcal{A}$, $\mathcal{A}^1$ and $\mathcal{A}^d$ in Figures 2.1, 3.1 and 3.4 respectively. $\mathcal{A}$ is non-deterministic at the location $S$ on $a$ when $x = 1$, since it has two edges, one to $S$ itself and other to $T$ which resets $y$. $\mathcal{A}^1$ focuses only on clock reduction and retains this non-determinism at location $S00$ on $a$ when $n = 1$ by having two edges one to $S00$ and other to $T10$. However, $\mathcal{A}^d$ is obtained by performing subset construction along with clock reduction. Thus in $\mathcal{A}^d$ the edge corresponding to the non-deterministic edges is $\{S00\} \xrightarrow{a, n=1?, n:=0} \{S11, T10\}$. We update the region component of $S00$ to $S11$ in the target state to reflect the difference between the values of $x$ in locations $S$ and $T$ in $\mathcal{A}$ after the edge. Hence, the edge $a, n = 0?$ from $\{S11, T10\}$ (due to $S11$) requires no time elapse as $a$ is valid from $S$ when $x = 1?$ (which is the value of $x$ in $S11$).

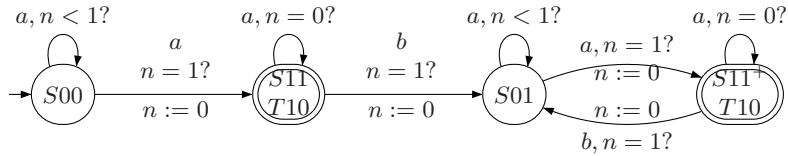

**Fig. 3.4.** Deterministic one clock IRTA $\mathcal{A}^d$ corresponding to the IRTA in Figure 2.1.

**Theorem 3.** *Let $\mathcal{A}$ be an IRTA and let $\mathcal{A}^d$ be the deterministic one clock IRTA constructed above. Then $L(\mathcal{A}) = L(\mathcal{A}^d)$.*

The proof is similar to the proof of Theorem 2 taking into consideration the subset construction.

**Complexity** From the definition of $\mathcal{A}^d$ given above, $L^d \subseteq 2^{L \times \mathcal{R}_I}$. Hence $|L^d| \leq 2^{|L| * |\mathcal{R}_I|} - 1 = 2^{|L| * (c_m + 2)^{|X|}} - 1$.

**Lemma 5.** *There is a non-deterministic IRTA $\mathcal{A}$ such that the smallest deterministic one clock IRTA $\mathcal{A}^1$ corresponding to it has exactly $2^{|L|} * (c_m+2)^{|X|} - 1$ locations, where $L$ is the set of locations of $\mathcal{A}$, $X$ is the set of clocks of $\mathcal{A}$ and $c_m$ is the maximum constant used in the guards of $\mathcal{A}$.*

*Proof.* See Appendix C. □

## 4 IRTA - Summary

We have given a simple and elegant technique to determinize the class IRTA and to reduce the number of clocks. The complexity bound we obtain is also optimal. If we allow $\epsilon$ moves in the IRTA $\mathcal{A}$, we can follow the clock reduction technique explained above by treating $\epsilon$ as a special symbol.

## 5 Integer resets in stopwatch automata

Stopwatches are variables whose rate of growth is either 0 or 1. Stopwatch automata (SWA) [8] obtained by adding stopwatches to timed automata render reachability undecidable while being expressively equivalent to linear hybrid automata [2]. Reachability is decidable for interrupt timed automata (ITA) [6], a variant of SWA with linear constraints, linear updates and restrictions on rates of growth and use of stopwatches in updates as well as constraints. To the best of our knowledge, this is the only known decidable variant of SWA. In this section, we explore the idea of integer resets in the context of stopwatch automata and define Integer Reset Stopwatch Automata (IRSA). We show that reachability is decidable for IRSA if diagonal constraints are not allowed. Further, in the absence of diagonal constraints, IRSA is determinizable, and closed under complementation, union and intersection. Undeciability of rechability of IRSA with diagonal constraints indicates that IRSA and ITA are incomparable.

An *integer reset stopwatch automaton* (IRSA) is a stopwatch automaton $\mathcal{A} = (L, L_0, \Sigma, X, Z, E, F, \eta)$ where (i) $L, L_0, F, X$ and $\Sigma$ are the same as in timed automata; (ii) $Z$ is a set of stopwatches; (iii) $\eta : L \to \{0,1\}^{|Z|}$ assigns the rate of growth of stopwatches in locations; (iv) $E \subseteq L \times L \times \Sigma \times C(X \cup Z) \times 2^{X \cup Z}$ is the set of transitions such that for every $e = (l, l', a, \varphi, \phi) \in E$, whenever $\phi \neq \emptyset$ or $\eta(l) \neq \eta(l')$, $\varphi$ consists of at least one atomic clock constraint of the form (a) $x = c$, for some $x \in X, c \in \mathbb{N}$, (b) $z = c$ for some $z \in Z, c \in \mathbb{N}$ provided $\eta(l)(z) = 1$.

The valuations of all variables is $\nu : X \cup Z \to \mathbb{T}$. Time elapse of $t$ units in a location $l \in L$, denoted as $\nu + t$ is as earlier (in Section 2) for clocks. For stopwatches it is defined as $\forall z \in Z, \nu + t(z)$ is $\nu(z) + t$ if $\eta(l)(z) = 1$, and is $\nu(z)$ if $\eta(l)(z) = 0$. Constraint satisfaction $\nu \models \varphi$ and resets $\nu[\phi := 0]$ are interpreted as defined earlier. It is easy to see that the semantics of IRSA are largely similar to those of timed automata. We follow the same notations as in Section 2.

**Proposition 1.** *Let $\mathcal{A}$ be an IRSA and $\nu$ be a valuation in any given run of $\mathcal{A}$. Then $\forall x, y \in X \cup Z, frac(\nu(x)) = frac(\nu(y))$.*

This proposition follows as a direct result of the definition of IRSA and Lemma 1. It allows us to consider $\mathcal{R}$ as the set of IRSA regions partitioning $\mathbb{T}^{X \cup Z}$. These are the same as IRTA regions (defined in Section 2) over the set $X \cup Z$.

Given an IRSA $\mathcal{A}$, we give a technique to convert it into a language equivalent IRTA $\mathcal{B}$. The construction is along the same lines as clock reduction in Section 3.1. We consider the locations of $\mathcal{B}$ to be $L \times \mathcal{R}_I$. Given a location $(l, \alpha)$ of $\mathcal{B}$, and a transition $(l, \alpha) \longrightarrow (l', \alpha')$, $\alpha$ is updated to $\alpha'$ on edges $l \longrightarrow l'$ of $\mathcal{A}$ that (i) reset a clock or stopwatch or (ii) $\eta(l) \neq \eta(l')$. For each stopwatch $z$ in $\mathcal{A}$, there is a clock $x_z$ in $\mathcal{B}$ simulating $z$. We consider atomic constraints in $\mathcal{A}$ to be of the form $x \in I$ where $x \in X \cup Z$ and $I \in \mathcal{I}$. For example, an edge with constraint $x = 2 \wedge z < 1$ can be represented as two edges with constraints $x \in [2] \wedge z \in [0]$ and $x \in [2] \wedge z \in (0,1)$ respectively. The formal construction of $\mathcal{B}$ from $\mathcal{A}$ is given below.

Given an IRSA $\mathcal{A} = (L, L_0, \Sigma, X, Z, E, F, \eta)$ construct an IRTA $\mathcal{B} = (L', L'_0, \Sigma, X \cup Z', E', F')$ as follows: (i) $L' \subseteq L \times \mathcal{R}_I$, $\mathcal{R}_I$ is the set of integral and saturated IRSA regions over $X \cup Z$; (ii) $L'_0 = L_0 \times \{\alpha_0\}$ where $\alpha_0 = ([0], [0], \cdots [0])$; (iii) $F' \subseteq F \times \mathcal{R}_I$; (iv) $Z'$ is a set of new clocks such that for every $z \in Z$, there is a unique clock $x_z$ in $Z'$ corresponding to $z$ via a bijection $Z' \leftrightarrow Z$; (v) $E' \subseteq L' \times \Sigma \times C(X \cup Z') \times 2^{X \cup Z'} \times L'$ is the set of transitions. A transition $(l, \alpha) \xrightarrow{a, \varphi', \phi'} (l', \alpha')$ is defined iff there exists a transition $l \xrightarrow{a, \varphi, \phi} l'$ in $E$ such that
(a) $\exists\ I \in \mathcal{I}$ such that $\alpha + I \models \varphi$. $\forall x \in X$, $(\alpha + I)(x) = \alpha(x) + I$ and $\forall z \in Z$, $(\alpha + I)(z) = \alpha(z) + I$ if $\eta(l)(z) = 1$, else $(\alpha + I)(z) = \alpha(z)$;
(b) $\varphi'$ is obtained by replacing $z \in c + \alpha(z)$ in $\varphi$ by $x_z \in c$, for all $z \in Z$;
(c) $\phi' = (\phi \cap X) \cup Z'$ if $\phi \neq \emptyset$ or $\eta(l) \neq \eta(l')$. Otherwise, $\phi' = \emptyset$;
(d) $\alpha' = (\alpha + I)[\phi := 0]$ if $\phi' \neq \emptyset$; else $\alpha' = \alpha$.

Each time a reset occurs or a rate changing edge is taken in $\mathcal{A}$, the corresponding edge in $\mathcal{B}$ resets all clocks in $Z'$ and updates $\alpha$ to contain the latest values of stopwatches. Hence constraints involving $Z'$ should pertain to the elapse since the last update of $\alpha$. Thus, the constraints in $\mathcal{B}$ replace $z \in c + \alpha(z)$ by $x_z \in c$. Appendix D gives an example of this construction and establishes that the resulting timed automaton is indeed an IRTA.

**Lemma 6.** *Let $\mathcal{A}$ be an IRSA and $\mathcal{B}$ be the IRTA constructed as above. Then $L(\mathcal{A}) = L(\mathcal{B})$.*

**Corollary 1.** *Reachability is decidable for the class IRSA. Further, it is closed under all boolean operations.*

Lemma 6 can be proved along the lines of Theorem 2. Corollary 1 follows from Lemma 6, Theorem 3 and decidability of emptiness of timed automata [3]. Note that the timed automaton $\mathcal{B}$ has at most $|L| \times (c_m + 2)^{|X \cup Z|}$ locations where $c_m$ is the maximum constant used in the constraints of $\mathcal{A}$. This bound can be proved to be tight employing the same technique as in Lemma 4.

**IRSA with diagonal constraints :** It is well known that diagonal constraints do not add to the expressive power of timed automata. However, we note that diagonal constraints involving stopwatches renders reachability undecidable for IRSA. It is easy to see that Minsky's two counter machine can be simulated using 3 stopwatches $x_1, x_2, x_3$ and one clock $g$ by following the encoding $c_1 = x_1 - x_2$ and $c_2 = x_2 - x_3$ for counters $c_1, c_2$. Incrementing $c_2$ is accomplished by a transition $\xrightarrow{g=0?}$ Ⓢ $\xrightarrow{g=1?g:=0}$ where $\eta(S)(x_3) = 0$ and $\eta(S)(x_i) = 1, \forall i < 3$. A simple diagonal constraint $x_1 - x_2 = 0?$ is sufficient to check if $c_1$ is zero.

Acknowledgement: We thank the anonymous reviewers for useful comments.

## Appendix

## A Equivalent runs in $\mathcal{A}$ and $\mathcal{A}^1$

Consider the IRTA $\mathcal{A}$ in Figure 2.1 and its corresponding one clock IRTA $\mathcal{A}^1$ in Figure 3.1. We now show a demonstration of the proof of Theorem 2 with an example. Recall that a state in $\mathcal{A}$ is of the form $(l_i, (\nu'_i(x), \nu'_i(y)))$ and a state in $\mathcal{A}^1$ is $(l_i, (\alpha_i(x), \alpha_i(y)), \mu'_i(n))$. We shall denote the clock intervals $[0], [1], (1, \infty)$ as $0, 1, 1^+$ respectively.

Consider a timed word $\rho = (a, 0.5)(a, 1)(a, 1)(b, 2)(a, 3)$. The run corresponding to $\rho$ in $\mathcal{A}$ is $r = (S, (0,0)) \xrightarrow{0.5} (S, (0.5, 0.5)) \xrightarrow{a, x \leq 1} (S, (0.5, 0.5)) \xrightarrow{1} (S, (1,1)) \xrightarrow{a, x \leq 1} (S, (1,1)) \xrightarrow{1} (S, (1,1)) \xrightarrow{a, x=1?, y:=0} (T, (1,0)) \xrightarrow{2} (T, (2,1)) \xrightarrow{b, y=1?, x:=0} (S, (0,1)) \xrightarrow{3} (S, (1,2)) \xrightarrow{a} (T, (1,0))$. There exists a run $r^1$ in $\mathcal{A}^1$ corresponding to $\rho$ given by $r^1 = (S, (0,0), 0) \xrightarrow{0.5} (S, (0,0), 0.5) \xrightarrow{a, n \leq 1} (S, (0,0), 0.5) \xrightarrow{1} (S, (0,0), 1) \xrightarrow{a, n \leq 1} (S, (0,0), 1) \xrightarrow{1} (S, (0,0), 1) \xrightarrow{a, n=1?, n:=0} (T, (1,0), 0) \xrightarrow{2} (T, (1,0), 1) \xrightarrow{b, n=1?, n:=0} (S, (0,1), 0) \xrightarrow{3} (S, (0,1), 1) \xrightarrow{a, n=1?, n:=0} (T, (1,0), 0)$. It is easy to see that $\nu'_i = \alpha_i + \mu'_i$ holds for all $i$.

## B Determinization of $\mathcal{A}^1$

In Section 3.1, we saw how to build a one clock possibly non-deterministic IRTA $\mathcal{A}^1$ for a given IRTA $\mathcal{A}$ with any number of clocks. As $\mathcal{A}^1$ is also an IRTA, we can apply the same technique outlined in Section 3.2 to obtain a deterministic one clock IRTA $\mathcal{A}^{1d}$. From Theorems 2 and 3, we know that $L(\mathcal{A}) = L(\mathcal{A}^1)$ and $L(\mathcal{A}^1) = L(\mathcal{A}^{1d})$. Hence $L(\mathcal{A}) = L(\mathcal{A}^{1d})$.

The Figure B.1 shows the deterministic one clock IRTA $\mathcal{A}^{1d}$ obtained from $\mathcal{A}^1$ in Figure 3.1 following definition in Section 3.2. Note that $\mathcal{A}^{1d}$ is the same as $\mathcal{A}^d$ in Figure 3.4.

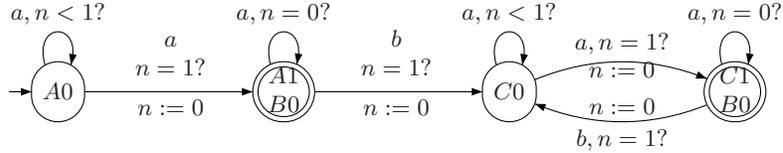

**Fig. B.1.** The deterministic one clock IRTA $\mathcal{A}^{1d}$ corresponding to the IRTA $\mathcal{A}^1$ in Figure 3.1. Here $A$, $B$ and $C$ represnt the locations $S00$, $T10$ and $S01$ of $\mathcal{A}^1$ respectively.

## C  Proof of Lemma 5

Consider the non-deterministic IRTA $\mathcal{A}$ in Figure C.1. It is clear that $\mathcal{A}^d$ in Figure C.1 has exactly $2^{|L|} * (c_m+2)^{|X|} - 1$ number of locations.

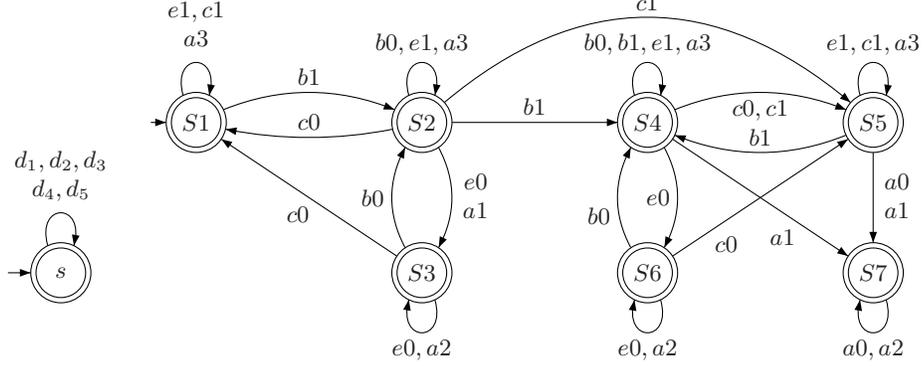

**Fig. C.1.** IRTA $\mathcal{A}$ and its deterministic IRTA $\mathcal{A}'$. The locations $S1$, $S2$, $S3$, $S4$, $S5$, $S6$ and $S7$ represent $\{S,0\}$, $\{(S,0),(S,1)\}$, $\{(S,1)\}$, $\{(S,0),(S,1),(S,1^+)\}$, $\{(S,0),(S,1^+)\}$, $\{(S,1),(S,1^+)\}$ and $\{(S,1^+)\}$ respectively. Here the symbols represent the following timed transitions $d_1 ::= b, x = 1?, x := 0$, $d_2 ::= b, x \geq 1?$, $d_3 ::= c, x = 1? x := 0$, $d_4 ::= c, x > 1?$, $d_5 ::= e, x \geq 1?$, $b0 ::= b, n = 0?$, $b1 ::= b, n = 1?, n := 0$, $c0 ::= c, n = 0?$, $c1 ::= c, n = 1?, n := 0$, $e0 ::= e, n = 0?$, $e1 ::= e, n = 1?$, $a0 ::= b, n = 0?; c, n = 0?; e, n = 0?$, $a1 ::= b, n \in (0,1)?; c, n \in (0,1)?; e, n \in (0,1)?$, $a2 ::= b, n > 0?; c, n > 0?; e, n > 0?$ and $a3 ::= b, n > 1?; c, n > 1?; e, n > 1?$.

The proof of Lemma 5 follows from the automaton $\mathcal{A}^d$ in Figure C.1.

## D  Details of Section 5

**$\mathcal{B}$ is an IRTA :** From the definition of $\mathcal{B}$, it is easy to observe the following.

- For every resetting edge $e$ in $\mathcal{A}$, there is a resetting edge $e'$ in $\mathcal{B}$ that resets all clocks in $Z'$ in addition to clocks mentioned in $e$.
- For every rate changing edge (source and target have different $\eta$ values) $e$ in $\mathcal{A}$, there exists an edge $e'$ in $\mathcal{B}$ which resets all clocks in $Z'$.

By definition of $\mathcal{A}$, we are assured that these kinds of edges occur at integer time units as they are accompanied by atomic constraints of the form (a) $x = c$, for some $x \in X, c \in \mathbb{N}$, (b) $z = c$ for some $z \in Z, c \in \mathbb{N}$ provided $\eta(l)(z) = 1$. Now consider the corresponding constraints in $\mathcal{B}$.

- If all the atomic constraints are over $X$, then they are the same in $\mathcal{B}$.
- If the atomic constraints in $\mathcal{A}$ involve $z = c$ (same as $z \in [c]$) then the corresponding constraint in $\mathcal{B}$ is of the form $x_z \in [c] - \alpha(z)$. As $\alpha \in \mathcal{R}_I$ over $X \cup Z$, $\alpha(z)$ is either integral or saturated. If $\alpha(z)$ is integral then $[c] - \alpha(z)$ is also integral. If $\alpha(z) = (c_m, \infty)$, then we are assured that there is no constraint of the form $z = c, c > c_m$ in $\mathcal{A}$ and hence no constraint $x_z \in [c] - \alpha(z)$ in $\mathcal{B}$.

From the above argument, it is clear that all resetting edges in $\mathcal{B}$ are accompanied by atomic constraints of the form $x \in [c], x \in X \cup Z'$. Thus, $\mathcal{B}$ is an IRTA.

### An IRSA $\mathcal{A}$ and its language equivalent IRTA $\mathcal{B}$

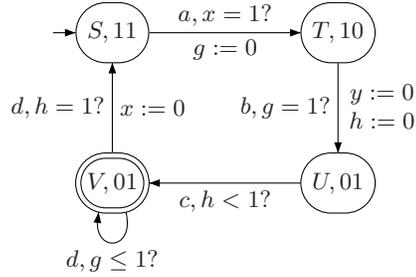

**Fig. D.1.** IRSA $\mathcal{A}$ with clocks $x, y$ and stopwatches $g, h$. The location $(T, 10)$ indicates that $\eta(T)(g) = 1$ and $\eta(T)(h) = 0$.

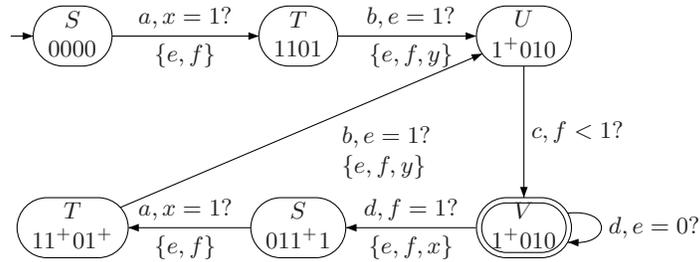

**Fig. D.2.** Timed automaton $\mathcal{B}$ which is language equivalent to IRSA in Figure D.1. Here the clock intervals $[0], [1], (1, \infty)$ are represented as $0, 1, 1^+$ respectively. Location $(T, 1101)$ stands for $(T, (x = 1, y = 1, g = 0, h = 1))$. The set of clocks to be reset is indicated on each edge. Clocks $e, f$ simulate the stopwatches $g, h$ respectively.

### Proof of Lemma 6 :

Language equivalence $L(\mathcal{A}) = L(\mathcal{A}^1)$ in Theorem 2 was established by proving that for a run in $\mathcal{A}$ there exists a run in $\mathcal{A}^1$ such that $\nu'_i = \alpha_i + \mu'_i$ always. A similar proof which inducts on the number of symbols in a timed word can be given for Lemma 6 too. The hypothesis is that for a state $(l_i, \nu_i')$ there exists a state $(l_i, \alpha_i, \mu_i')$ in $\mathcal{B}$ such that $\nu_i' \cap X = \mu_i' \cap X$ and $\forall z \in Z, \nu_i'(z) \in \alpha_i(z) + \langle \mu_i'(x_z) \rangle$. Thus, $\nu_i'(z) \models z \in c + \alpha_i(z)$ iff $\mu_i'(x_z) \models x_z \in c$ for all $i$.